\documentclass[twocolumn]{IEEEtran}
\pdfoutput=1
\ifCLASSINFOpdf
   \usepackage[pdftex]{graphicx}
  \DeclareGraphicsExtensions{.pdf,.jpeg,.png}
\else
   \usepackage[dvips]{graphicx}
   \DeclareGraphicsExtensions{.eps}
\fi

\usepackage{amssymb, amsmath, amsfonts, amsthm}
\usepackage{dsfont, bbm}
\usepackage{float, graphics, xspace, graphicx}
\usepackage[usenames,dvipsnames]{color}
\usepackage{cite, epsfig, tikz, standalone}
\usepackage{algorithm,algorithmicx,algpseudocode}
\usepackage{subfigure}

\usepackage[absolute]{textpos}
\graphicspath{{./figs/}}

\makeatletter
\theoremstyle{plain}

\let\oldnl\nl
\newcommand{\nonl}{\renewcommand{\nl}{\let\nl\oldnl}}

\def\twon #1{\left\|#1\right\|_2}

\def\atomn #1{\left\|#1\right\|_{\cA}}

\def\abs #1{\left|#1\right|}

\def\st{\text{subject to }}

\def\bC{\mathbb{C}}

\def\m #1{\boldsymbol{#1}}

\def\cA{\mathcal{A}}

\def\cM{\mathcal{M}}

\def \bee{\begin{equation}}
\def \ene{\end{equation}}

\def\beq{\begin{eqnarray}}
\def\enq{\end{eqnarray}}

\def\equ #1{\begin{equation}#1\end{equation}}

\def\sbra #1{\left(#1\right)}

\def\lbra #1{\left\{#1\right\}}

\def\st {\text{ s.t. }}

\newcommand{\FRF}{\mathbf{F}_\mathrm{RF}}
\newcommand{\FRFH}{\mathbf{F}^{\rm H}_{\mathrm{RF}}}
\newcommand{\PBB}{\mathbf{P}_\mathrm{BB}}

\newcommand{\w}{\mathbf{w}}

\newcommand{\f}{\mathbf{f}}

\newcommand{\Hk}{\mathbf{H}_k}
\newcommand{\HkH}{\mathbf{H}_k^{\rm H}}
\newcommand{\HjH}{\mathbf{H}_j^{\rm H}}

\newcommand{\hvk}{\mathbf{h}_{{\rm v}}}
\newcommand{\hv}{\mathbf{h}_{{\rm v}}}

\newcommand{\Ak}{\mathbf{A}}

\newcommand{\mb}[1]{{\mathbf #1}}

\newcommand{\Rmnum}[1]{\expandafter\@slowromancap\romannumeral #1@}

\begin{document}


\title{MmWave Channel Estimation via Atomic Norm Minimization for Multi-User Hybrid Precoding}
\author{
    \IEEEauthorblockN{Junquan Deng\IEEEauthorrefmark{1}, Olav Tirkkonen\IEEEauthorrefmark{1} and Christoph Studer\IEEEauthorrefmark{2}}\\
    \IEEEauthorblockA{\IEEEauthorrefmark{1}Department of Communications and Networking, Aalto University, Finland \\
                      \IEEEauthorrefmark{2}School of Electrical and Computer Engineering, Cornell University, NY, USA \vspace*{-4mm}
    }
}
\maketitle
\begin{abstract}
To perform multi-user multiple-input and multiple-output transmission in millimeter-wave~(mmWave) cellular systems, the high-dimensional channels need to be estimated for  designing the multi-user precoder. Conventional grid-based Compressed Sensing~(CS) methods for mmWave channel estimation suffer from the basis mismatch problem, which prevents accurate channel reconstruction and degrades the precoding performance. This paper formulates mmWave channel estimation as an Atomic Norm Minimization~(ANM) problem. In contrast to grid-based CS methods which use discrete dictionaries, ANM uses a continuous dictionary for representing the mmWave channel. We consider a continuous dictionary based on sub-sampling in the antenna domain via a small number of radio frequency chains. We show that mmWave channel estimation using ANM can be formulated as a Semidefinite Programming~(SDP) problem, and the  channel can be accurately estimated via off-the-shelf SDP solvers in polynomial time. Simulation results indicate that ANM can achieve much better estimation accuracy compared to grid-based CS, and significantly improves the spectral efficiency provided by multi-user precoding.
\end{abstract}


\section{Introduction}
In future millimeter-wave~(mmWave) cellular networks, large antenna arrays are expected to be applied at the Base Station~(BS) to serve multiple User Equipments~(UE) in dense urban scenarios. To perform Multi-User Multiple-input and Multiple-Output~(MU-MIMO) hybrid precoding~\cite{Alkhateeb2014, AlkhateebFeedback2015}, accurate Channel State Information~(CSI) is necessary. As mmWave channels typically compromise a few strong propagation paths, Compressed Sensing~(CS) methods have been considered for mmWave channel estimation~\cite{alkhateeb2015comp,AlkhateebFeedback2015,Two_stages2016,Alkhateeb2014,Lee2016}. In order to apply CS to channel estimation, a discretization procedure is generally adopted to reduce the continuous angular or delay spaces to a finite set of grid points. Assuming that the Direction of Departure~(DoD) and Direction of Arrival~(DoA) lie on the grid, the virtual channel representation is sparse with few non-zero entries. The channel estimation problem then can be addressed with a specific measurement matrix and a recovery algorithm, e.g., the Orthogonal Matching Pursuit~(OMP)~\cite{Lee2016}. However, as the actual signals are continuous and will not fall on the discrete points, basis mismatch will degrade recovery performance~\cite{offthegrid}. Although finer grids can reduce the reconstruction error, they require more computation resources.

In the literature, most of the multi-user hybrid precoding schemes assume perfect CSI at the transmitters~(see e.g., \cite{Hybrid_2016,Alternating2016}). However, grid-based CS for channel estimation will lead to the basis mismatch problem and hence channel estimation errors, which will degrade the performance of MU-MIMO precoding such as Zero-Forcing~(ZF)~\cite{AlkhateebFeedback2015}. Recently, a continuous basis pursuit technique with auxiliary interpolation points~\cite{CBP_2011} was used for mmWave channel estimation in~\cite{CBP_ICCW2017}, which shows that adding interpolation points to the original grids can improve channel estimation accuracy considerably.
This paper considers a continuous dictionary for multi-user mmWave channel estimation in order to completely eliminate the basis mismatch error. We proposed a continuous dictionary based on antenna-domain sub-sampling via a much smaller number of Radio Frequency~(RF) chains in the hybrid architecture. The sparse mmWave channel estimation is formulated as an Atomic Norm Minimization (ANM) problem~\cite{ANM_Bhaskar_2013}, which can be solved via Semidefinite Programming~(SDP) in polynomial time. ANM has been considered for massive MIMO channel estimation ~\cite{AND_ICC_2015, MmWANMICC2017}, wherein a fully-digital architecture is assumed which has much higher complexity compared to the hybrid architectures. To show the advantage of ANM based on antenna-domain sub-sampling in the hybrid architecture, we investigate a low-complexity MU-MIMO hybrid precoding scheme using the estimated channel information. Finally, we evaluate the performance of ANM and the precoding scheme via simulations in realistic scenarios. Simulation results regarding the channel estimation error and user spectral efficiency are provided, which verify  the efficacy of our solutions.


\section{System model}\label{sec:model}
\subsection{Architectures of BS and UE}
\begin{figure}[t]
\begin{center}
\includegraphics[width=8.5cm]{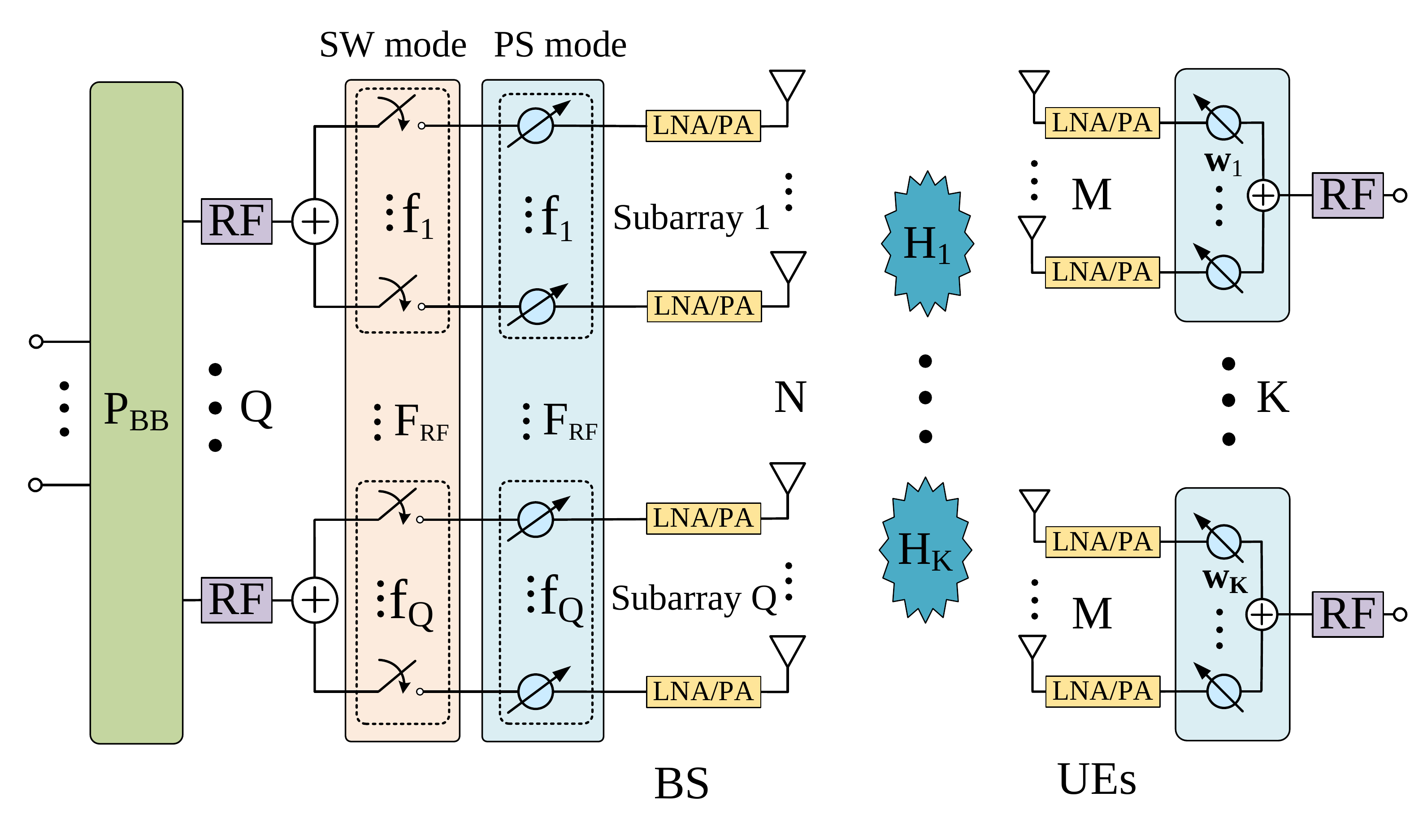}
\end{center}
\vspace*{-4mm}
\caption{System architectures for  BS and UE, the RF part of BS has two modes of operation: 1) phase-shifter-based mode~(PS mode) for directional data transmission and 2) switch-based mode~(SW mode) for channel estimation.}
\label{fig:Arc}
\end{figure}

We consider a mmWave cellular system where the BS has ${N}$ antennas and $Q$ RF chains to serve $K = Q$ UEs in downlink~(DL). In contrast to \cite{Hybrid_2016,alkhateeb2015comp}, a BS architecture  equipped with both Phase Shifters~(PS) and switches is proposed, as shown in Fig.~\ref{fig:Arc}. The BS antennas are grouped into $Q$ sub-arrays. Antennas in each sub-array are associated with one particular RF chain. With the proposed architecture, the BS can adopt two working modes: 1) the PS mode with a partially-connected PS network~(as A2 in~\cite{Hybrid_2016}); 2) the SW mode with a partially-connected switch network~(as A6 in~\cite{Hybrid_2016}).
When working in
SW mode, the BS contains a fully-digital architecture with $Q$
antennas. The
switch network enables accessing
the received signal from an individual antenna, which is useful for
channel estimation, especially when the Signal-to-Noise
Ratio~($\rm{SNR}$) is high. The cost of adding dedicated switches to
the phase-shifter network is moderate as the implementation complexity
of switches is typically lower than that of phase
shifters~\cite{Hybrid_2016}.
In contrast, the PS network is used to perform analog beamforming and combining for both beam training and data transmission.

At the UE side, each UE is assumed to have a single RF chain with $M$ antennas and $M$ phase shifters. Such UEs are able to perform single-data-stream analog beamforming.
This UE architecture has
low hardware complexity and
energy consumption, making it suitable for mobile UEs~\cite{rappaport2014millimeter}.

The BS has two kinds of RF precoder/combiner depending on the working modes. In SW mode, the sub-array RF precoder/combiner $\f_q$ is an antenna selection vector with one non-zero element. In PS mode, $\f_q$ is a sub-array steering vector. For UEs, the RF precoder/combiner $\mb{w}_k$ of UE $k$ is a beam steering vector. We assume that BS and UE have $Q_{\rm{ps}}$-bits quantized PSs~\cite{Beamforming_Design_2016}. The number of beams that can be steered by BS or UE depends on the PS resolution $Q_{\rm ps}$ rather than the number of antennas, and is equal to $2^{Q_{\rm ps}}$. The coefficient set available for each PS is denoted by $\mathbb{P} = \{\omega^0,\omega^1,\ldots,\omega^{2^{Q_{\rm ps}} - 1}\}$ where $\omega \!=\! \mathrm{e}^{j2\pi/{2^{Q_{\rm ps}}}}$ is the minimum angular domain separation.
The quantized PSs can adjust the phases of the signals and have constant modulus. Specifically, the RF beamforming codewords are
\begin{equation}\label{eqxx}
\f_q = {\mathbf{1}_{\Omega_q}} \!\odot\! \m{f}_{\omega}(c,N),\,\,\,
\w_k = \m{f}_{\omega}(c,M),
\end{equation}
where $\Omega_q$ is the antenna index set for $q$th sub-array, ${\mathbf{1}_{\Omega_q}}$ is a binary valued vector with $|\Omega_q|$ ones indexed by $\Omega_q$, $\m{f}_{\omega}(c, i) = [1,\omega^c,\omega^{2c},\ldots,\omega^{(i-1)c}]^{T}$ with $c \in \{0,1,\ldots,2^{Q_{\rm ps}}\!-\!1\}$.
\subsection{Channel Model}
MmWave channels are highly directional and contain a few dominant paths in the angular domain~\cite{rappaport2014millimeter}. Due to the extremely short wave-length, diffraction effects in mmWave frequencies are small; only Line-of-Sight~(LoS) and reflected paths are significant. As a consequence, the number of paths is smaller than the number of BS antennas. Suppressing the vertical directivity of the channel, for the $k$th UE, the DL channel is modeled as~\cite{alkhateeb2015comp},
\begin{equation}\label{channelmodel}
\Hk=\sum\nolimits_{l=1}^{L}{  \alpha_{l}  \mathbf{a}_{\rm{UE}}(\theta_{l})  \mathbf{a}_{\rm{BS}}(\phi_{l})^{\rm H} } \in \mathbb{C}^{M\times N},
\end{equation}
where $L$ represents the number of paths, and $\alpha_{l}$ denotes the complex gain of the $l$th path. In addition, $\mathbf{a}_{\rm{BS}}(\phi_{l})$ and $\mathbf{a}_{\rm{UE}}(\theta_{l})$ represent the BS and UE array response vectors for the $l$th path, where $\phi_{l}$ is the DoD at BS, and $\theta_{l}$ is DoA at UE.
Considering plane-wave model and Uniform Linear Array~(ULA) for both BS and UE, the array steering vectors can be written as
\begin{equation}\label{aBSaUE}
\begin{aligned}
\mb a_{\rm BS}(\phi)\!  &=\![1, \mathrm{e}^{j \frac{2\pi}{\lambda}d\sin(\phi)},\ldots,\mathrm{e}^{j(\!N-1\!)\frac{2\pi}{\lambda}d\sin(\phi)} ]^{\rm T},\,\,\,\\
\mb a_{\rm UE}(\theta)\!&=\![1, \mathrm{e}^{j \frac{2\pi}{\lambda}d\sin(\theta)},\ldots,\mathrm{e}^{j(\!M-1\!)\frac{2\pi}{\lambda}d\sin(\theta)} ]^{\rm T},
\end{aligned}
\end{equation}
where $\lambda$ is carrier wavelength and the antenna spacing here is assumed to be $d=\lambda/2$. We assume that the system is working in the Time Division Duplex~(TDD) mode and the DL and uplink~(UL) channels are reciprocal.  Denote the BS baseband combiner or precoder as $\mathbf{P_{\rm{BB}}}\!=\![\mb p_1,\ldots,\mb p_K] \!\in\! {\mathbb{C}}^{ Q \times K}$,
the BS RF combiner or precoder
as $\mathbf{F_{\rm{RF}}} = [\f_1,\!\ldots\!, \f_Q ] \in {\mathbb{C}}^{ N \times Q}$,
and the RF combiner or precoder for all $K$ UEs as
${\mb W} = [\mb w_1,\ldots,\mb w_K]\in {\mathbb{C}}^{M \times K}$.
The DL and the UL received signals including interference and noise for UE $k$ are
\begin{equation}\label{eql}
\begin{aligned}
z_{{\rm u}} &=\mb w_k^{\rm H} {\Hk} \FRF \left(\mb p_k x_k \!+\! \sum\nolimits_{j\neq k} \!\mb p_j x_j\right)+\mb w_k^{\rm H}\mb {n}_{{\rm u}},\\
z_{{\rm b}} &=\mb p^{\rm H}_{k} \FRFH \left(\HkH \mb w_k y_k\!+\! \sum\nolimits_{j\neq k}\!\!\HjH \mb w_j y_j+\mb n_{{\rm b}}\right),
\end{aligned}
\end{equation}
where $x_k$ and $y_k$ are the DL and UL signals, satisfying
${\mathbb{E}(x_k^{} x_k^*) = \frac{{\rho}_{\rm{BS}}}{K}}$ and $ \mathbb{E}(y_k^{} y_k^*) = \frac{\rho_{\rm{UE}}}{M}$ with transmit powers as $\rho_{\rm BS}$  and $\rho_{\rm UE}$. In addition, $\mb n_{{\rm u}} \in \mathbb{C}^{M \times 1} $ is the $\mathcal{C}\mathcal{N} ( 0,\sigma^{2}_{\rm u})$ UE noise vector with noise power $\sigma^{2}_{\rm u}$ per antenna, and $\mb n_{{\rm b}} \in \mathbb{C}^{N \times 1} $ is the $\mathcal{C}\mathcal{N} ( 0,\sigma^{2}_{\rm b})$ BS noise vector with noise power $\sigma^{2}_{\rm b}$ per antenna. For DL transmission, ${{\bf{F}}} = \mathbf{F_{\rm{RF}}} \mathbf{P_{\rm{BB}}}$ satisfies power constraint ${{\rm{tr}}( {{\bf{F}}{{\bf{F}}^{\rm H}}} ) \le 1}$. For UL transmission, $\mb w_k$ satisfies $\|\mb w_k\|_2^2\leq 1$.

\section{Problem formulation for Channel Estimation}\label{sec:OMP}
For channel training in UL, we assume that the $k$th UE transmits a pilot sequence $\mb s_k \in \mathbb{C}^{1 \times T_{\rm s}}$ using UE beam $\mathbf{w}_k$. Assuming $\mathbf{P_{\rm{BB}}} = \mathbf{I}_Q$, the received training signal at BS is
\begin{equation}
\begin{aligned}
\mathbf{Y} 
           &=  \FRFH (\sum\nolimits_{k=1}^{K}{\mathbf{h}}_k \mathbf{s}_{k} + \mathbf{N}_{\rm b}).\\
\end{aligned}
\label{ULtraining}
\end{equation}
Here $\mb h_k = \HkH \w_{k}$ is the effective  channel for UE $k$. The user pilots are assumed to satisfy
$\mathbf{s}_{k}\mathbf{s}_{k'}^{\rm H} =  T_{\rm s} \rho_{\rm UE} \delta_{k,k'}$, with $\delta_{i,j}$
the Kronecker delta function. Correlating $\mathbf{Y} $ with  $\mb s_k$, we get
\begin{equation}
\begin{aligned}
\mathbf{y}_k &= \mathbf{Y} \mathbf{s}^{\rm H}_k=  {T_{\rm s} \rho_{\rm UE} } \FRFH  {\mathbf{h}}_k  + \FRFH \mathbf{N}_{\rm b} \mathbf{s}^{\rm H}_k.
\end{aligned}
\label{ULtraining2}
\end{equation}
Assuming that the BS takes $T$ snapshots of measurements with different combining matrices, stacking the $ W = T \times Q$ measurement samples gives
\begin{equation}
\mb z \!=\!
\begin{bmatrix}
\mb y_{1,k} \\
\mb y_{2,k} \\
\vdots \\
\mb y_{T,k} \\
\end{bmatrix}
 \!=\!
\underbrace{
T_{\rm s} \rho_{\rm UE}
\begin{bmatrix}
{\mb F}_{1,\rm RF}^{\rm H}  \\
{\mb F}_{2,\rm RF}^{\rm H}  \\
\vdots \\
{\mb F}_{T,\rm RF}^{\rm H}  \\
\end{bmatrix}
}_{\mb \Phi}
{\mathbf{h}}_k  +
\underbrace{
\begin{bmatrix}
{\mb F}_{1,\rm RF}^{\rm H}  \mathbf{N}_{1,\rm b} \\
{\mb F}_{2,\rm RF}^{\rm H}  \mathbf{N}_{2,\rm b}\\
\vdots \\
{\mb F}_{T,\rm RF}^{\rm H}  \mathbf{N}_{T,\rm b}\\
\end{bmatrix}
\mathbf{s}^*_k
}_{\mb n}
,
\label{zk2}
\end{equation}
where $\mb \Phi$ is the sensing matrix and ${\mb n}$ is the noise after combining.
The objective of BS channel estimation is to estimate the effective channel. From~\eqref{channelmodel}, we have
\begin{equation}\label{hk}
{\mathbf{h}}_k=\HkH \w_{k}=\sum\nolimits_{l=1}^{L}{ {\beta}_{l} \mathbf{a}_{\rm{BS}}(\phi_{l})},
\end{equation}
where ${\beta}_{l} = (\alpha_{l} (\mathbf{w}_k)^{\rm H} \mathbf{a}_{\rm{UE}}(\theta_{l}))^*\!$. The effective channel ${\mathbf{h}}_k$ contains fewer significant paths compared to the full MIMO channel $\mb H_k$. To estimate ${\mathbf{h}}_k$, one can estimate DoAs $\{\phi_{l}\}$ and the complex gains $\{{\beta}_{l}\} $ for the paths and then reconstruct the channel. In grid-based CS methods, a discrete dictionary ${\mb \Psi_{\rm BS}} \!= \!\left[  \mb a_{\rm BS}(\phi_1),\ldots,\mb a_{\rm BS}(\phi_{G_{\rm b}}) \right]$ with $G_{\rm b}$ bases is used to represent the channel $\mb h_k$. We consider a grid in which
$\{\sin(\phi_1),\ldots,\sin(\phi_{G_{\rm b}})\}$ are evenly distributed in $(-1,1]$~\cite{Lee2016}. The channel vector is represented as
\begin{equation}
{\mathbf{h}}_k = {\mb \Psi_{\rm BS}} \hvk,
\end{equation}
where $ \hvk \in\mathbb{C}^{ G_{\rm b} \times 1}$ is the sparse virtual channel in dictionary ${\mb \Psi_{\rm BS}}$. Denoting  $\mathbf{A} =$ $ \mb \Phi {\mb \Psi_{\rm BS}} $, to estimate the sparse virtual channel $\hvk $, one can formulate the following optimization problem:
\begin{equation}\label{CSmin2}
\begin{aligned}
&\underset{\hvk}{\mathrm{minimize}} \left\Vert \hvk \right\Vert_0
&\mathrm{s.t.}  && \| \mb z_k - \mathbf{A} \hvk \|_{2}^2 \leq \eta.
\end{aligned}
\end{equation}
where $\eta$ is an optimization parameter and typically set as the noise power after combining. In contrast to~\cite{Lee2016}, we consider a switch-based method for constructing the sensing matrix, in which only $Q$ individual antennas are sampled during one measurement  snapshot. We call this method Antenna Domain Sub-Sampling~(ADSS). The ADSS  sensing matrix has the following structure:
\begin{equation}
\begin{aligned}
\mb \Phi
           &={T_{\rm s} \rho_{\rm UE} }  [\mb e_{i_{1,1}},\mb e_{i_{1,2}}, \ldots, \mb e_{i_{1,Q}}, \ldots, \mb e_{i_{T,Q}}]^{\rm H},
\end{aligned}
\label{sensing2}
\end{equation}
where the unit vector $\mb e_{i_{t,q}}$ has one entry equal to 1 at position $i_{t,q}$, and $i_{t,q}$ is the antenna index for the $q$th RF chain in the $t$th measurement snapshot. Furthermore, if we define the antenna index set $\Omega = \{i_{1,1}, i_{1,2}, \ldots, i_{1,Q}, \ldots, i_{T,Q}\}$ for all sampled antennas, we have
\begin{equation}
\mb z = \mb \Phi \mb h_k + \mb n = {T_{\rm s}\rho_{\rm UE}}   (\mb h_k)_{\Omega} + \mb n.
\label{eq:zksimple}
\end{equation}
A variety of algorithms have been proposed for obtaining an
approximate solution to~\eqref{CSmin2} at polynomial complexity. OMP
is a preferred method due to its simplicity and fast implementation~\cite{Lee2016}. To solve~\eqref{CSmin2} based on the measurement data $\mb z $
in~\eqref{eq:zksimple}, we use OMP as in Algorithm~\ref{alg:OMP}. The
computational complexity of OMP is proportional to the number $G_{\rm
  b}$ of grid points. Step 3 in Algorithm~\ref{alg:OMP} needs
$\mathcal{O}(WG_{\rm b})$ computations and solving the Least
Square~(LS) problem in step 5 requires $\mathcal{O}(W
|\mathcal{I}_t|^2 )$.

The joint design of the sensing matrix $\mb \Phi$ and dictionary
matrix $\mb \Psi$ plays a key role in achieving accurate channel
estimation. To reconstruct $\hvk$, $\Ak$ should have low
mutual coherence $\mu(\Ak)$, which
is the largest normalized inner product for two different columns of
$\Ak$. Specifically, when the path directions lie on the defined grid, it can be shown that OMP can recover $\hvk$ in the
noiseless case if~\cite{Hybrid_2016}
\begin{equation}
\mu(\Ak) = \underset{i\neq j}{\mathrm{max}} \, \frac{|\mb{a}_{i}^{\rm H} \, \mb{a}_{j}|}{\|\mb{a}_{i}\|_{\rm 2} \cdot  \|\mb{a}_{j}\|_{\rm 2}} < \frac{1}{2{L}-1},
\label{eq:muAk}
\end{equation}
where $\mb{a}_{i}$ and $\mb{a}_{j}$ are two different columns of
$\Ak$. The lower bound of $\mu(\Ak)$ is given by the Welch bound~\cite{Hybrid_2016}, which
decrease as the number of measurement $W$ increases.
\begin{algorithm}[!t]
\textbf{Input}: The sensing matrix $\mb \Phi$, the dictionary matrix ${\mb \Psi_{\rm BS}}$ and $\mathbf{A} =$ $ \mb \Phi {\mb \Psi_{\rm BS}} = [{\mb a}_1, {\mb a}_2,\ldots,{\mb a}_{G_{\rm b}}]$, the measurement vector $\mathbf{z}$, and a threshold $\delta$.
\begin{algorithmic}[1]
\State Iteration counter $t\gets0$, basis vector set $\mathcal{I}_t \gets \emptyset$, virtual channel $\hv \gets \m 0 \in \mathbb{C}^{G_{\rm b}\times 1}$, residual error $\mathbf{r}_t\gets\mathbf{z}$
\While{$\|\mathbf{r}\| > \delta$ and $t < G_{\rm b}$}
\State $g^{\star} \gets \arg\max_g \|{\mb a}^{\rm H}_g \, \mathbf{r}_t \|_{2}$ \Comment{Find new basis vector}
\State $t \gets t+1$, $\mathcal{I}_t \gets \mathcal{I}_{t-1} \bigcup\,\{g^{\star}\}$ \Comment{Update vector set}
\State $\m{x}^{\star} \gets \arg\min_{\m{x}} \|\mathbf{z} \!-\! (\mathbf{A})_{\mathcal{I}_{t}} \, \m{x}\|_{2}$ \Comment{LS method}
\State $\mathbf{r}_t \gets \mathbf{z} - (\mathbf{A})_{\mathcal{I}_{t}} \, \m{x}^{\star}$ \Comment{Update residual error}
\State $(\hv)_{\mathcal{I}_t} \gets \m{x}^{\star} $ \Comment{ Update the virtual channel}
\EndWhile
\end{algorithmic}
\textbf{Output}: Estimated channel $\hat{\mathbf{h}} = {\mb \Psi_{\rm BS}} \hv$
\caption{OMP for effective channel estimation}
\label{alg:OMP}
\end{algorithm}

\section{Atomic Norm Minimization for Equivalent Channel Estimation}\label{sec:ANM}
When using the dictionary ${\mb \Psi_{\rm BS}}$ for channel estimation
via OMP, the estimated path angles are considered to be on a discrete
grid $\{\phi_1,\ldots,\phi_{G_{\rm b}}\}$, which introduces the basis
mismatch problem~\cite{offthegrid}. To avoid this, we consider
a continuous dictionary for estimating the effective channel.

To estimate the effective channel based on UL measurements, the
following constrained optimization problem can be formulated:
\equ{\underset{\mb{h}}{\mathrm{minimize}}\, \cM
  \sbra{\mb{h}}\,\,\,
  \st \|\mb{z} - \mb{\Phi}\mb{h}\|_2^2\leq \eta.
  \label{eq:sparseoptinz1}}
Here $\cM(\mb{h})$ denotes a sparse
metric to be minimized for a channel vector $\mb{h}$. The noise is
assumed to be bounded by $\|\mb n\|^2_2 \leq \eta $, and $\mb{h}$ can
be treated as the signal of interest which needs to be estimated based
on the observed data $\mb{z}$ in the context of line spectral
estimation~\cite{ANM_Bhaskar_2013}.
Problem~\eqref{eq:sparseoptinz1} differs from~\eqref{CSmin2} as it is
formulated without a discrete dictionary. Instead
of~\eqref{eq:sparseoptinz1}, one can solve a regularized optimization
problem as \equ{\underset{\mb{h}}{\mathrm{minimize}} \,\, \xi
  \cM\sbra{\mb{h}} + \frac{1}{2}\twon{\mb{z} - \mb \Phi\mb{h}}^2,
  \label{eq:sparseoptinz2}} where $\xi > 0$ is a regularization
parameter related to $\eta$. There are different choices for $\cM(\mb
h)$~\cite{offthegrid}. Here, we consider the atomic norm $\cM(\mb h)
=\atomn{\mb{h}} $ proposed in~\cite{offthegrid}. The atomic norm
has been  used in grid-less compressive sensing for a range of
applications~\cite{offthegrid,ANM_Bhaskar_2013,AND_ICC_2015,MmWANMICC2017},
including DoA estimation, line spectral estimation and massive MIMO
channel estimation. The continuous set of atoms used to represent
$\mb{h}$ is defined as \equ{\cA= \lbra{\mb{a}_{\rm BS}\sbra{\mb
      \phi}\alpha: \; \phi\!\in\!
    \left(\!-\frac{\pi}{2},\!\frac{\pi}{2}\right],\alpha\in\bC,
  \abs{\alpha}=1}.} The atomic norm is the gauge
  function~\cite{ExactFreq2016} of the convex hull
  $\text{conv}\sbra{\cA}$. Formally, the atomic norm can be written as
  \equ{
\begin{split}
\atomn{\mb{h}}
&=\! \inf \lbra{\, g>0: \mb{h}\in g \cdot \text{conv} \sbra{\cA}}\\
&=\! \inf \left\{\sum\nolimits_{i} b_i: \mb{h} = b_{i}\sum\nolimits_i \mb{a}_i, b_i>0, \mb{a}_i \in \cA \right\}. \\
\end{split} \label{eq:atomn}
}
The atomic norm is a continuous counterpart of the $\ell_1$-norm used in on-grid CS methods.
It can be computed efficiently via semidefinite programming~\cite{offthegrid}. Specifically, $\|\mb{h}\|_{\cA}$ defined in \eqref{eq:atomn} is the optimal value of the following matrix trace minimization problem:
\equ{
\underset{\m{u},t}{\mathrm{minimize}} \,\, \frac{1}{2}(t+u_1)\,\,\, \st \begin{bmatrix} \mathcal{T}(\m{u}) & \mb{h} \\ \mb{h}^{\rm H} & t \end{bmatrix}\succeq \m{0}, \label{thm:AN_SDP}
}
where $\mathcal{T}(\m{u})$ is a Hermitian Toeplitz matrix with the first row as $\m{u} = [u_1,\ldots,u_N]^{\rm T}$.
In the noisy case, using the atomic norm, we can rewrite the optimization problem~\eqref{eq:sparseoptinz2} as
\begin{equation}
\begin{aligned}
\underset{\mb D \succeq \m{0} }
{\mathrm{minimize}} &\, \frac{\xi}{2}(t+u_1)\! +\! \frac{1}{2}\twon{\mb{z}-\mb\Phi\mb{h}}^2 \,\,\,\,\\
& \st \mb D = \begin{bmatrix} \mathcal{T}(\m{u}) &\! \mb{h} \\ \mb{h}^{\rm H} &\! t \end{bmatrix}.
\label{formu:AN_SDP4}
\end{aligned}
\end{equation}
The above problem is a SDP which can be solved by off-the-shelf convex optimization tools in polynomial time~\cite{sdpt3}. The computational complexity is $\mathcal{O}\left((N+1)^6\right)$ in each iteration using the interior point method.

To guarantee that $\mb{h}$ can be exactly recovered via ANM, the measurement matrix $\mb \Phi$ and the number of atoms in $\mb h$~(i.e., the number of paths $L$ in the user channel) should satisfy a condition which is similar to the mutual coherence condition in on-grid CS. To define this condition for ANM, a concept called spark for the continuous set of atoms is introduced. Let us define the transformed continuous dictionary based on the measurement matrix $\mb \Phi$ as
\begin{equation}
\cA_{\mb \Phi} = \left\{\mb \Phi \mb{a}_{\rm BS}(\phi): \phi \in \left(-\frac{\pi}{2},\frac{\pi}{2}\right] \right\}.
\end{equation}
Similar to the definition of matrix spark, $\mathrm{spark}(\cA_{\mb \Phi})$ is defined as the smallest number of atoms which are linearly dependent in $\cA_{\mb \Phi}$. From~\cite{ExactFreq2016}, the problem~\eqref{eq:sparseoptinz1} with $\eta = 0$ in the noiseless setting has a unique solution if the number of paths satisfies
\begin{equation}
L < \frac{\mathrm{spark}(\cA_{\mb \Phi})}{2}.
\label{eq:spark}
\end{equation}
Note that the spark in~\eqref{eq:muAk} satisfies $\mathrm{spark}(\Ak) \geq 1 + \left(\mu(\Ak)\right)^{-1}$, so the  condition above  is equivalent to~\eqref{eq:muAk} in the case of discrete atoms. Thus the $\mathrm{spark}(\cA_{\mb \Phi})$ should be as large as possible so that ANM can recover a channel with as many paths as possible.
For a specific BS array, one may notice that $\mathrm{spark}(\cA_{\mb \Phi})$ depends only on $\mb \Phi$. For ADSS via switches, $\mathrm{spark}(\cA_{\mb \Phi})$ depends on the number of sampled antennas $|\Omega|$ and the sampled antenna indexes. It was shown in~\cite{ExactFreq2016} that $ 2 \leq \mathrm{spark}(\cA_{\mb \Phi}) \leq |\Omega|+1 $, and a random $\Omega$ or a $\Omega$ with $|\Omega|$ consecutive integers can achieve a sufficiently large spark.


\section{Multi-user Downlink Precoding/Combining  based on the Estimated Channels}\label{sec:MU_pre}
In this section, we consider the designs of downlink BS precoder and UE combiners. To design the BS precoder $\{\FRF,\PBB\}$ and UE combiners $\mb W$ optimally, full CSI $\mathcal{H} = \{\mb H_1, \mb H_2,\ldots,\mb H_K\}$ needs to be known. However, acquiring $\mathcal{H}$ would require each UE to send huge amount of pilots which consumes lots of resources. A sub-optimal method, which works well when user channels are sparse, is that each UE first finds its best beam which aligns with the most significant path, then UEs transmit their pilots to BS for multi-user channel estimation as discussed in sections~\ref{sec:OMP} and \ref{sec:ANM}. As a result, the BS only needs to estimate the user effective channel $\mb h_k$ which has much fewer coefficients and significant paths to be estimated. The BS will then design the its MU-MIMO precoder based on those estimated effective channels. For UEs, the beam codebook is
\begin{equation}
\mathcal{U} \!=\! \left\{ \sqrt{1/M}\m{f}_{\omega}(c,M): c \!\in\! \{0,\ldots,2^{Q_{\rm ps}}\!-\!1\}\! \right\}.
\end{equation}
For BS, the DL training beam codebook used by the $q$th sub-array is
\begin{equation}
\mathcal{F}_q \!=\! \left\{ {\mathbf{1}_{\Omega_q}} \!\odot\! \m{f}_{\omega}(c,N): c \!\in\! \{0,\ldots,2^{Q_{\rm ps}}\!-\!1\}\! \right\}.
\end{equation}
Sub-array $q$ will use one beam from $\mathcal{F}_q$ for UE beam training.

\begin{figure}
\centering
\subfigure[BS sub-array]{\includegraphics[width=4cm,angle = 0]{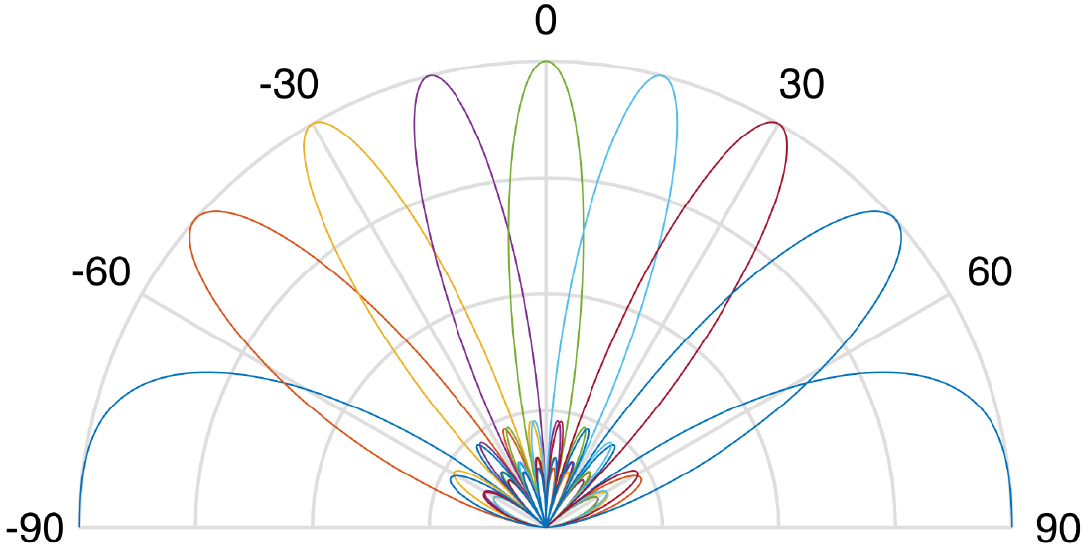}}\,\,\,\,
\subfigure[UE array]{\includegraphics[width=4cm,angle = 0]{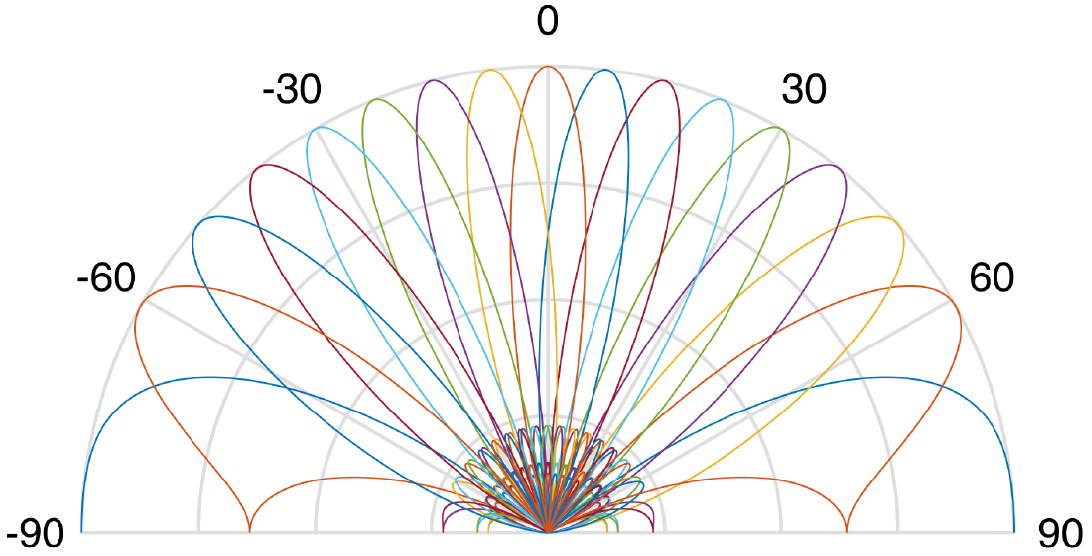}}
\caption{(a) Designed $Q$ DL sub-array beams with $N=64,Q=8, Q_{\rm ps} = 4 $; (b) Designed 16 UE beams with $M=8, Q_{\rm ps} = 4$.}
\label{DLBSUEBeams}
\end{figure}

The objective of the UE beam training is to find a UE beam which maximizes the norm of the effective channel
\begin{equation}
\begin{aligned}
&\underset{{\w_k} \in \mathcal{U}}{\mathrm{maximize}} \,  \|{\mathbf{w}}_k^{\rm H} \Hk \|^2_{\rm 2},
\end{aligned}
\label{opt:MaxUEbeam}
\end{equation}
The problem in~\eqref{opt:MaxUEbeam} is equivalent to finding the best steering vector ${\w_k} \in \mathcal{U}$ that aligns with the paths that have large path gains. In principle, to find the best UE codeword, BS needs to transmit DL training beams which cover all the paths. However, this would increase the training overhead.
We propose using all sub-arrays to transmit $Q$ DL pilots with $Q$ sub-array beams simultaneously which are evenly distributed in $[-\pi/2,\pi/2 )$. Fig.~\ref{DLBSUEBeams}~(a) shows the designed DL beams used for the BS sub-arrays and Fig.~\ref{DLBSUEBeams}~(b) shows the UE beam codebook we used.
The MU-MIMO precoding scheme based on the UE beam training and the effective channel estimation is summarised and shown in Algorithm~\ref{alg:Hybrid}. Noticing that DoAs and DoDs of the propagation paths change much slower than the channel coefficients~\cite{rappaport2014millimeter}, the best UE beam remains the same during a period which is much longer than channel coherence time, and the UE beam tracking can be performed less frequently.
\begin{algorithm}[!ht]
\begin{algorithmic}[1]
\State BS transmits $Q$ DL pilots using $Q$ sub-arrays
\State Each UE $k\in \{1,2,\ldots,K\}$ receives DL pilots using codewords from UE codebook $\mathcal{U}=\{\mathbf{w}^{(1)},\ldots,\mathbf{w}^{(P)}\}$ and calculate the received power with the $p$th UE beam and $q$th BS DL beam, each UE then has a received power matrix $\mathbf{P}_k = [\mb p_1, \mb p_2, \ldots,\mb p_Q]\in \mathbb{R}^{P\times Q}$. Let $\mb c\leftarrow \mb 0_{P \times 1}$,
\For {$q = 1,2,\ldots,Q$}
\State Find the best UE beam $i^{\star}$ for the $q$th BS beam:\\
\quad\,\, $i^{\star} \leftarrow {\rm{argmax}}_i \, (\mb p_q)_i$,
$(\mb c)_{i^{\star}} \leftarrow (\mb c)_{i^{\star}} + (\mb p_q)_{i^{\star}}$.
\State Find the final best UE beam:\\
\quad\,\, $p^{\star} \leftarrow {\rm{argmax}}_i \, (\mb c)_i$,
$\mathbf{w}_k \leftarrow \mathbf{w}^{(p^{\star})}$
\EndFor
\State UE $k$ transmits UL pilot to BS using $\mathbf{w}_k$, the BS estimates the effective user channel $\mb h_k$ using OMP or ANM methods as described in sections~\ref{sec:OMP} and~\ref{sec:ANM}; the BS has an estimated multi-user channel $\hat{\mb H} = [\hat{\mb h}_1, \hat{\mb h}_2, \ldots, \hat{\mb h}_K ]^{\rm H}$. Let $\FRF = [\f_1,\!\ldots\!, \f_Q ] \gets \mb 0_{N\times Q}$,
\For {$q \in \{1,2,\ldots,Q\}$}
\State Select UE $k = q$, find $\f_q =\mathrm{argmax}_{{\f} \in \mathcal{F}_q } |\hat{\mb{h}}_k^{\rm H} \f|^2$
\EndFor
\State Design $\PBB$ as $\PBB = \frac{(\hat{\mb H}\FRF\!)^{\!-\!1}}{\|\FRF (\hat{\mb H}\FRF\!)^{\!-\!1}\|_{\rm F}}$ using ZF.
\end{algorithmic}
\caption{Joint UE RF combining and BS MU-MIMO precoding}
\label{alg:Hybrid}
\end{algorithm}

\section{Numerical Results}\label{sec:Sim}
\begin{figure}[t]
\begin{center}
\includegraphics[width=7.5cm]{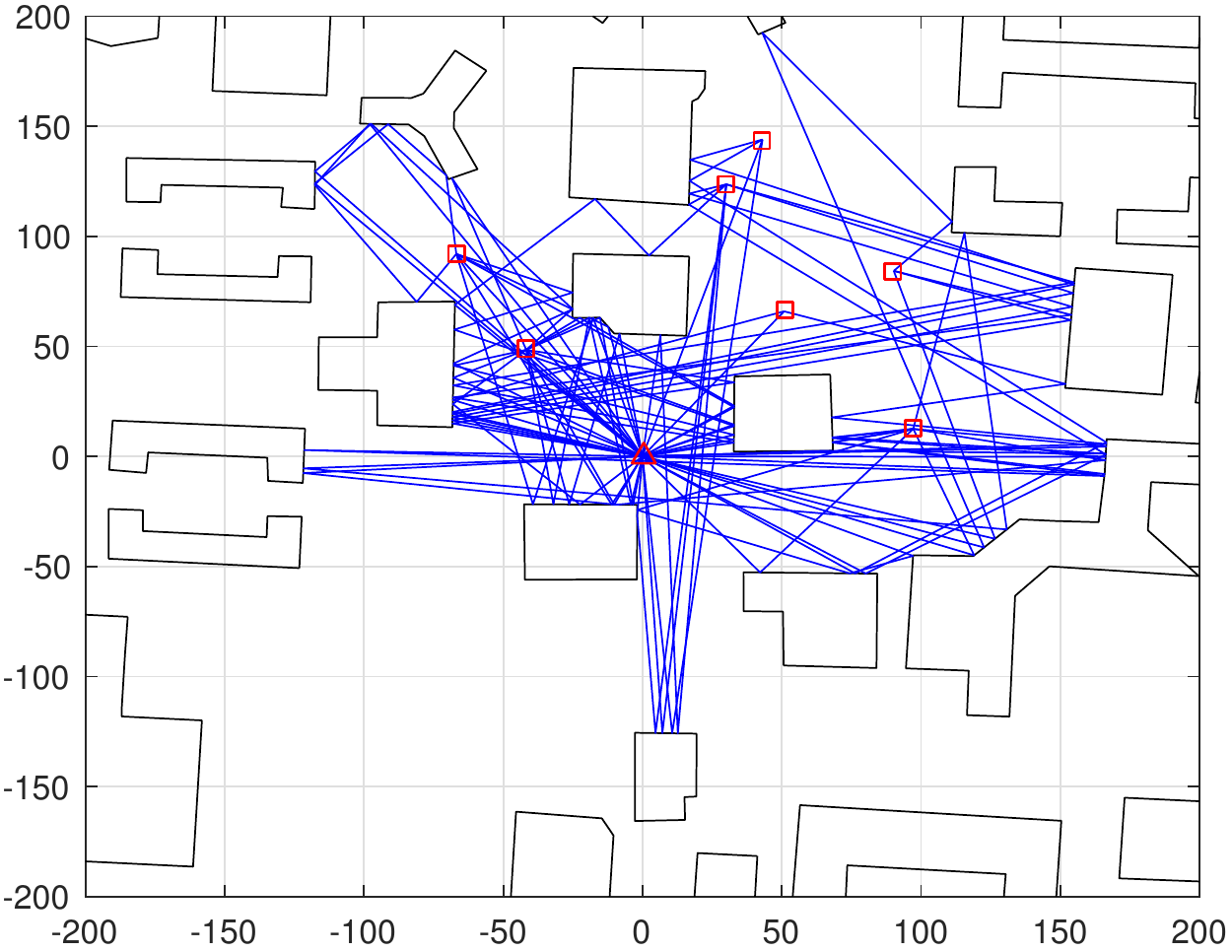}
\end{center}
\vspace*{-4mm}
\caption{Simulation scenario in a 400$\times$400 $\text{m}^2$ area, the polygons represent the buildings with walls, red triangle is the BS and red squares are the UEs. Ray-tracing is used to generate the propagation paths which are shown in blue lines.}
\label{Scenario}
\end{figure}

In this section, the performance of the proposed channel estimation method and the precoding scheme are evaluated via numerical simulations. We assume $M=8$ for UEs, and $N=64$, $Q=8$ for BS. The BS antennas are grouped into $Q$ sub-arrays with equal size. We consider an outdoor small cell where BS is lower than buildings, and signals will be reflected or blocked by the walls. A ray-tracing channel model\cite{28GHzRayTracing2015} is considered to generate the channels. The reflection coefficients are computed based on the Fresnel equation and up to 3rd order reflections are taken into account. The relative permittivities of building walls are uniformly distributed between 3 and 7. The BS is at the center while UEs are dropped outdoor with distances to BS smaller than 150 meters, as shown in Fig.~\ref{Scenario}. Blocked UEs which cannot find a LOS or a reflected path are not considered in the simulation.
The path coefficient in~\eqref{channelmodel} is
\begin{equation}
\alpha_{l} = \mathrm{e}^{j\psi}\sqrt{G_0 x^{-2} g_{1}(\theta_{l}) g_{2}(\phi_{l})\prod\nolimits_{i=0}^R |r_i|^2},
\end{equation}
where $G_0 = 10^{-6.14}$ is the omnidirectional path gain~\cite{mmWEvaluation2014} at reference distance  one meter, $x$ is the path propagation distance in meter, $\psi \sim U(0,2\pi)$ is a random phase, $g_{1}(\theta_{l})$ and $g_{2}(\phi_{l})$ are UE and BS antenna element patterns, respectively, $R$ is the reflection order and $r_i$ is the $i$th reflection coefficient. For a LOS path, we have $R=0,\,r_0 = 1$. More details of simulation parameters are given in Tabe~\ref{table1}.

\begin{table}[t]
\centering
\small
\caption{Simulation Parameters}
\begin{tabular}{c|c|c}\hline
\textbf{ Parameter} & \textbf{Symbol} & \textbf{value} \\ \hline
Carrier frequency & $f_c$    &  28 GHz      \\ \hline
System bandwidth  & $b_{w}$  &  256 MHz     \\ \hline
OFDM subcarrier number & $N_c$     & 256 \\ \hline
UE total Tx power   & $ N_c \!\times\! \rho_{\rm UE} $ & 23 dBm  \\ \hline
BS total Tx power   & $ N_c \!\times\! \rho_{\rm BS} $ & 40 dBm  \\ \hline
UE noise power &    $N_c \!\times\! \sigma^{2}_{\rm u}$ & -89 dBm \\ \hline
BS noise power &    $N_c \!\times\! \sigma^{2}_{\rm b}$ & -86 dBm \\ \hline
UE antenna pattern  &   $g_{1}(\theta) $   & omnidirectional \\  \hline
BS antenna pattern  &   $g_{2}(\phi) $ & defined in~\cite{3GPP900}      \\ \hline
Phase-shifter resolution  & $Q_{\rm ps} $ & 4 bits \\ \hline
Pilot sequences      &  $\mb s_k$  &  Zadoff-Chu, $T_s = 63$ \\ \hline
\end{tabular}
\label{table1}
\end{table}

\subsection{Effective User Channel Estimation via OMP and ANM}

\begin{figure}[!t]
\begin{center}
\includegraphics[width=8cm]{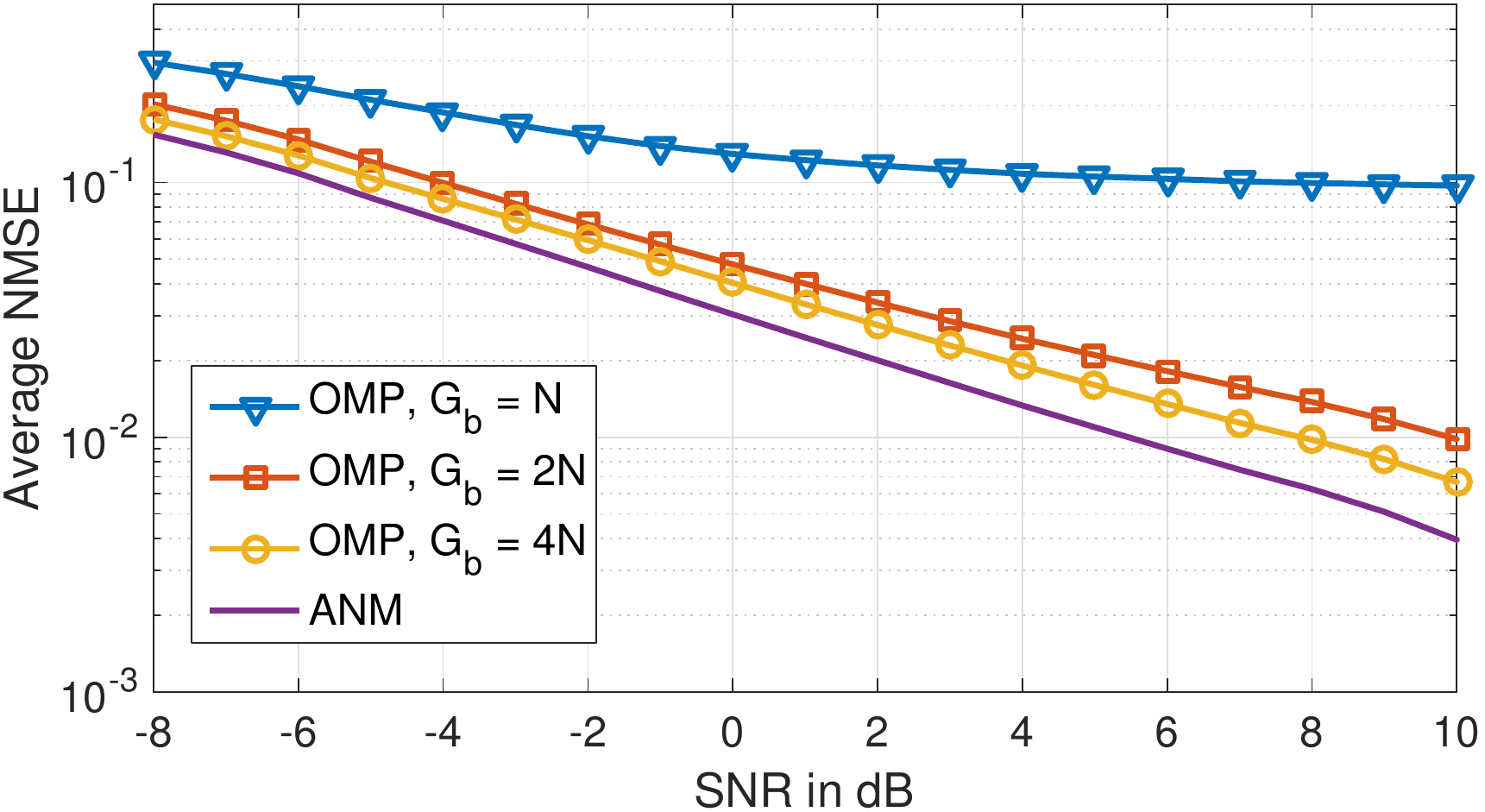}
\end{center}
\vspace*{-4mm}
\caption{The average channel normalized estimation error for different channel estimation schemes.}
\label{NMSE}
\end{figure}

All UEs transmit orthogonal pilots to BS using their best beams, and BS receives them via ADSS. In these simulations, $T=2$ measurement snapshots are used, so we have $T\times Q = 16$ antennas sampled pseudo-randomly via the switch network.
The ANM problem~\eqref{formu:AN_SDP4} is solved via the $\rm SDPT3$~\cite{sdpt3} solver.
The OMP algorithm applys a non-redundant dictionary with $G_b\!=\!N$
and redundant dictionaries with $G_b\!=\! 2N,\, 4N $.
We estimate channel estimation performance, in terms
of the Normalized Mean Square Error~(NMSE) $\mathbb{E}\{ {\|\mb h_k
  \!-\! \hat{\mb h}_k\|^2_{\rm 2}}/{\|\mb h_k\|^2_{\rm 2}}\}$.
In total 8000 UEs are simulated. Results are presented as a function of
$\mathrm{SNR} = {\rho_{\rm UE}\|\Hk\|^2_{\rm F}}/{(MN\sigma^2_{\rm b})}$.
As illustrated in Fig.~\ref{NMSE}, increasing $G_{\rm b}$
directly results in better estimation accuracy for OMP. The ANM method outperforms
OMP since it avoids basis mismatch by using the continuous dictionary.
As $G_{\rm b}$ increases, the gap between OMP and ANM decreases at the
cost of increasing computational complexity.
When a non-redundant dictionary with $G_b = N$ is used for
OMP, OMP suffers severely from basis mismatch;
an error floor develops in the hign SNR regime.
In comparison, ANM achieves much better estimation accuracy,
 especially in the high SNR regime, effectively removing the error floor.
Inaccurate channel estimation degrades the interference mitigation performance in MU-MIMO precoding techniques. As a result, ANM would be a strong candidate for the hybrid precoding solution discussed in Section~\ref{sec:MU_pre}. We also found that in the simulated scenario, the user effective channel $\mb h_k$ is sparse in angular domain. In most cases, there are one to three significant paths even when the original channel $\mb H_k$ has much more paths. This ensures that the condition~\eqref{eq:spark} is satisfied and ANM can recover the channel with high probability. In the case that the condition~\eqref{eq:spark} is not satisfied, ANM aims to recover those strongest paths and moderate channel estimation accuracy is attainable. Compared to OMP, solving the ANM problem using $\rm SDPT3$ entails higher complexity. In practice, an efficient solver based on the Alternating Direction Method of Multipliers~(ADMM)~\cite{Chi_2016} technique could be adopted to solve the ANM problem.

\subsection{Spectral Efficiency Performance of Multi-User Precoding}
\begin{figure}[t]
\begin{center}
\includegraphics[width=8cm]{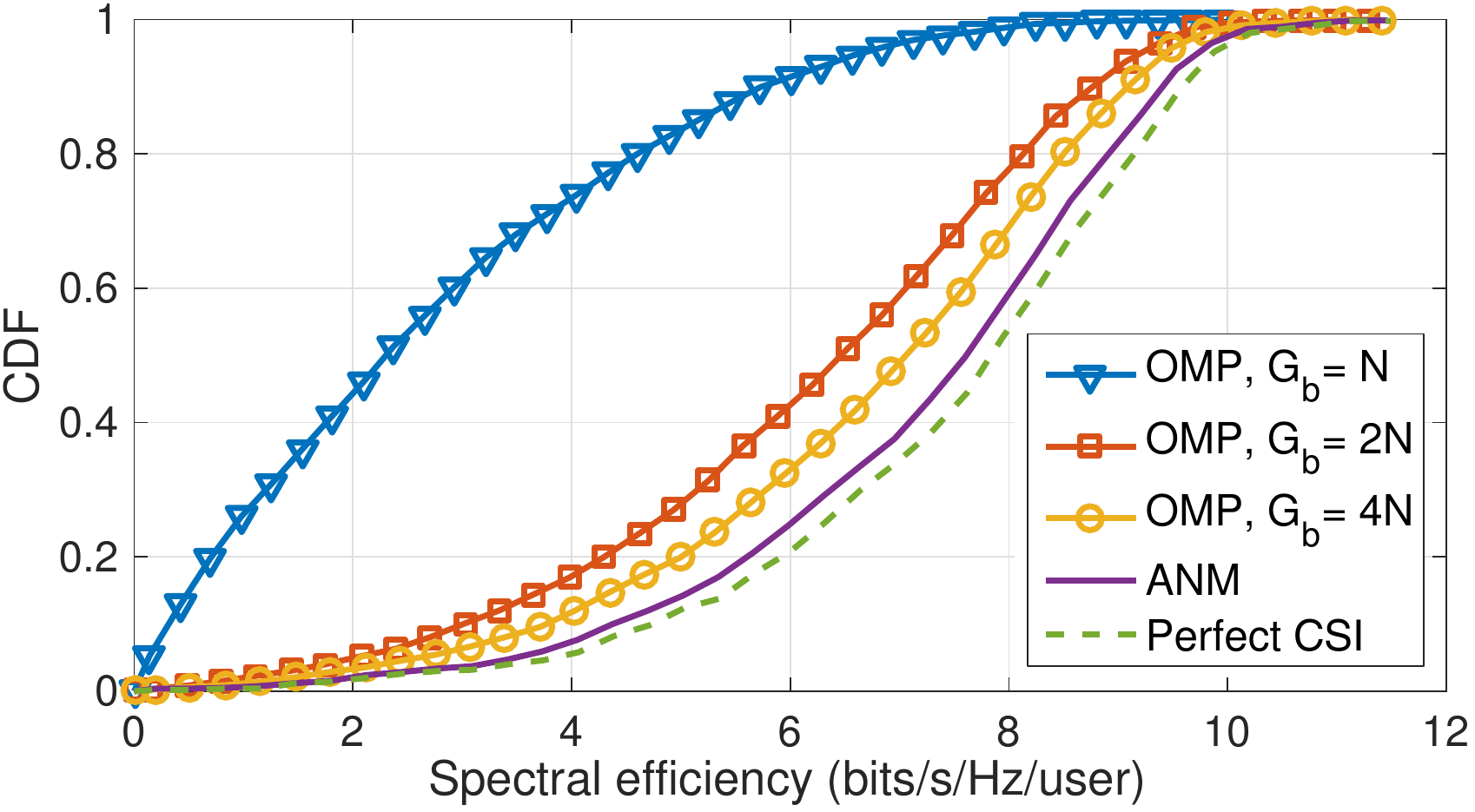}
\end{center}
\caption{Performance of multi-user hybrid precoding with sub-arrays, with different channel estimates.}
\label{CDF2}
\end{figure}

To evaluate the multi-user precoding performance, 8000 UEs are randomly divided into 1000 groups and 8 UEs in each group are served simultaneously.
The Cumulative Distribution Function~(CDF) for user spectral efficiency is collected, where SE is estimated as $\log_2(1+\mathrm{\gamma})$
from the Shannon formula with $\gamma$ the signal-to-interference-plus-noise ratio. As shown in Fig.~\ref{CDF2}, hybrid precoding with ANM-based channel estimation can provide much better performance compared to an OMP-based scheme since it provides accurate channel estimates for the effective user channel. The baseband ZF precoding is sensitive to channel estimation error and the error will damage the user orthogonality provided by ZF. On average, OMP with $G_b = N$, $2N$ and $4N$ loses 63\%, 18\% and 11\% of the spectral efficiency with perfect CSI.  By comparison, ANM loses only 3.4\%, which verify the efficacy of ANM.

\section{Conclusion}\label{sec:concl}
This paper has proposed a low-complexity hybrid architecture in which an inexpensive switch network is added to the subarrays to facilitate channel estimation. We have formulated the mmWave channel estimation as an ANM problem which can be solved via SDP with polynomial complexity. Compared to the grid-based CS methods, the proposed ANM approach uses a continuous dictionary with sub-sampling in antenna domain via the switch network. Simulation results show that ANM can achieve much better channel estimation accuracy compared to grid-based CS methods, and can help MU-MIMO hybrid precoding to achieve significantly better user spectral efficiency performance.

\bibliographystyle{IEEEtran}
\bibliography{IEEEabrv,ref}
\end{document}